\documentclass[twoside,twocolumn,english,pra,aps,superscriptaddress]{revtex4-1}
\usepackage[T1]{fontenc}
\usepackage{color}
\usepackage{babel}
\usepackage{textcomp}
\usepackage{amsmath}
\usepackage{amssymb,amsfonts,mathrsfs}
\usepackage{amsthm}
\usepackage{graphicx}
\usepackage[unicode=true,pdfusetitle,
 bookmarks=true,bookmarksnumbered=false,bookmarksopen=false,
 breaklinks=false,pdfborder={0 0 0},backref=false,colorlinks=true]
 {hyperref}
\usepackage{breakurl}

\makeatletter
\theoremstyle{plain}

\makeatother

\begin{document}

\preprint{This line only printed with preprint option}

\title{Quantum algorithm for evaluating operator size with Bell measurements}

\author{Xi-Dan Hu}
\affiliation{Guangdong Provincial Key Laboratory of Quantum Engineering and Quantum Materials, School of Physics and Telecommunication
Engineering, South China Normal University, Guangzhou 510006, China}

\author{Tong Luo}
\affiliation{Guangdong Provincial Key Laboratory of Quantum Engineering and Quantum Materials, School of Physics and Telecommunication Engineering, South China Normal University, Guangzhou 510006, China}


\author{Dan-Bo Zhang}
\email{dbzhang@m.scnu.edu.cn}
\affiliation{Guangdong-Hong Kong Joint Laboratory of Quantum Matter, Frontier Research Institute for Physics, South China Normal University, Guangzhou 510006, China}
\affiliation{Guangdong Provincial Key Laboratory of Quantum Engineering and Quantum Materials, School of Physics and Telecommunication
Engineering, South China Normal University, Guangzhou 510006, China}

\date{\today}

\begin{abstract}
Operator size growth describes the scrambling of operators in quantum dynamics and stands out as an essential physical  concept for characterizing quantum chaos. Important as it is, a scheme for direct measuring operator size on a quantum computer is still absent. Here, we propose a quantum algorithm for direct measuring the operator size and its distribution based on Bell measurement. The algorithm is verified with spin chains and meanwhile, the effects of Trotterization error and quantum noise are analyzed. It is revealed that saturation of operator size growth can be due to quantum chaos itself or be a consequence of quantum noises, which make a distinction between quantum integrable and chaotic systems difficulty on noisy quantum processors. Nevertheless, it is found that the error mitigation will effectively reduce the influence of noise, so as to restore the distinguishability of quantum chaotic systems. Our work provides a feasible protocol for investigating quantum chaos on noisy quantum computers by measuring operator size growth.
\end{abstract}

\maketitle

\section{Introduction}
Characterizing quantum chaos has drawn intensive interests in recent years due to its fundamental role in understanding quantum statistical mechanics~\cite{MarkSrednicki1994,ABohrdt2017,AmosChan2018}. One important perspective to study quantum chaos is to investigate the quantum information scrambling~\cite{BrianSwingle2016,AavishkarA2017,SilviaPappalardi2018,TibraAli2020,TianruiXu2020,ManojKJoshi2020,XMi2021}, a concept to describe how  local information can be scrambled into the whole system and become nonlocal. The information scrambling can be quantified by the out-of-time-ordered correlations (OTOCs)~\cite{SHShenker2014a,SHShenker2014b,JunLi2017,ChengJuLin2018,CWvonKeyserlingk2018,WenLeiZhao2021}. Remarkably, recent rapid advances in quantum processors enable us to observe information scrambling by measuring OTOCs, which is unusual as typical correlation functions are time-ordered.

An alternative approach to understand information scrambling is to directly investigate the operator spreading~\cite{PHosur2016,DARoberts2018,AdamNahum2018,VedikaKhemani2018,SarangGopalakrishnan2018,XLQi2019a,XLQi2019b,SanjayMoudgalya2019,DEParker2019,SKZhao2021,ThomasSchuster2022,EwanMcCulloch2022}. In this picture, a local operator after evolution can spread into a linear combination of highly nonlocal operators, and the number of nonlocal operators can be exponentially enlarged, making extraction of information encoded in the initial local operator impractical. In other words, the operator size can grow with time until reaching the system size. The operator size growth can be intuitive for characterizing and understanding quantum chaotic systems. Remarkably, for open-system dynamics where OTOCs can have difficulty telling information scrambling of the evolution of the system itself apart from noises of the environment, the operator size distribution can still give a faithful characterization. In this regard, operator size growth provides a promising avenue to study information scrambling other than OTOCs. However, there is still a lack of feasible schemes to directly measure the operator size on quantum processors, except for indirect measurement by inferring from data obtained by quantum quenches from an ensemble of random initial states~\cite{XLQi2019b,ManojKJoshi2020}.

In this paper, we propose a quantum algorithm that can directly measure the operator size and its distribution, which is feasible on near-term quantum computers. The scheme is based on a mapping between Pauli operators and Bell states. By preparing a product of Bell states, the operator of Heisenberg evolution will imprint the information of the product of Pauli operators into Bell states, and by Bell measurements, the operator size and its distribution can be extracted. In the numeral simulation, we consider both Trotterization error and quantum noises that are related to implementation. For spin chains,
it is found that quantum noises may make distinguishing between integrable and chaotic systems difficult, as both the operator sizes grow to saturation. Nevertheless, we show that the feature of operator size oscillation can be restored for the integrable system by error mitigation~\cite{KristanTemme2017,SuguruEndo2018,ManpreetSinghJattana2020,ArmandsStrikis2021,ElizabethRBennewitz2022,YasunariSuzuki2022}. Our work points out that the characterization of quantum chaos by measuring the operator size can be feasible on near-term quantum devices with error mitigation.

The paper is organized as follows. In Sec.~\ref{s2} we introduce the definition of operator size and propose a quantum algorithm based on Bell measurements to evaluate the operator size. Then in Sec.~\ref{s3} we present numeral simulation results for spin chains and analyze the effects of Trotterization, quantum noises, and error mitigation. Finally, we make conclusions in Sec.~\ref{s4}.

\section{Evaluating operator size by Bell measurements}
\label{s2}
In this section, we first give a definition of operator size for
quantum systems. Then, we illustrate how Bell measurements~\cite{SenruiChen2022,HsinYuanHuang2022} can be exploited to measure the operator size.
\subsection{Operator size}
To be concrete without loss of generality, we consider a lattice model where each site is a qubit~(spin-half). The Hamiltonian can be written as a summation of local terms, $H=\sum_i \lambda_iH_i$, each local term $H_i$ a product of Pauli operators. In the Heisenberg picture, the time evolution of a quantum system can be encapsulated into the time-dependent operator $\hat{O}(t)$, which satisfies the motion equation,
\begin{equation}
\frac{\partial \hat{O}(t)}{\partial t}=i[\hat{H},\hat{O}(t)].
\end{equation}
The solution is $\hat{O}(t)=e^{i\hat{H}t}\hat{O}(0)e^{-i\hat{H}t}$.

The time evolution of quantum systems can be very complicated, even if the Hamiltonian itself is simple and the initial operator $\hat{O}(0)$ is a single Pauli operator. To see this, the operator $\hat{O}(t)$ can be written as
\begin{equation}\label{eq:expand_O}
\hat{O}(t)=\left[\sum_{n=0}^{\infty}\frac{(i\hat{H}t)^n}{n!}\right]\hat{O}(0)\left[\sum_{n=0}^{\infty}\frac{(-i\hat{H}t)^n}{n!}\right].
\end{equation}
For a generic quantum many-body system, a single Pauli operator can evolve into a linear combination of products of Pauli operators, which may involve many ones by expanding $H^n$ in Eq.~\eqref{eq:expand_O}.

For a generic N-qubit lattice system, the operator $\hat{O}(t)$ can be written under the Pauli basis~(product of Pauli operators),
\begin{equation}\label{OPt} 
	\begin{split}
		\hat{O}(t)=\sum_{\mathbf{k}=0}^{4^N-1}C_{\mathbf{k}}(t)\hat{P}_{\mathbf{k}},~~~~ \hat{P}_{\mathbf{k}}=\bigotimes^N_{n=1}\sigma^{k_n}_n.
	\end{split}
\end{equation}
Here we have used a quaternary number $\mathbf{k}=k_1k_2...k_N$( $k_n=0, 1, 2, 3$) to label all Pauli basis in order and Pauli matrices are $\sigma^0=I$, $\sigma^1=X$, $\sigma^2=Y$, $\sigma^3=Z$. The coefficient can be evaluated as,
\begin{equation}\label{Ckt}
C_{\mathbf{k}}(t)=2^{-N}\text{tr}(\hat{O}(t)\hat{P}_{\mathbf{k}}).
\end{equation}
Once the initial operator $\hat{O}(0)$ is Hermitian, it can be  seen that $\hat{O}(t)$ is Hermitian and all coefficients $C_{\mathbf{k}}(t)$ are real numbers. Thus, it requires $4^N$ real numbers to fully characterize the operator $\hat{O}(t)$, which becomes inaccessible for a generic quantum many-body system due to the exponential growth. Nevertheless, one may study some properties of $\hat{O}(t)$ that are of physical interest.

From the aspect of quantum information scrambling, one remarkable feature of $\hat{O}(t)$ is to investigate how the operator size is growing with time evolution. For quantum chaos, the initial local information can be scrambled into the whole system. The information is distributed extensively in the system and thus becomes nonlocal. The Pauli basis is suitable to study the degree of extensive by defining its operator size, which counts how many non-I operators are in $\hat{P}_{\mathbf{k}}$. Initially, the operator size $\hat{O}(0)$ as a Pauli operator on a site is one. With the time evolution, the operator $\hat{O}(t)$ is a superposition of Pauli basis with different operator sizes, and it is necessary to use an averaged operator size. For quantum chaos, the averaged operator size grows with time until saturation, which is $O(L)$.

Let us write explicitly the averaged operator size, which is
\begin{equation}\label{OS} 
L[\hat{O}(t)]\equiv\sum_{\mathbf{k}=0}^{4^N-1}\left|C_{\mathbf{k}}(t)\right|^2\times l_{\hat{P}_{\mathbf{k}}}.
\end{equation}
Here $l_{\hat{P}_{\mathbf{k}}}$ is the operator size of Pauli basis $\hat{P}_{\mathbf{k}}$ defined as
\begin{equation}\label{lP}
l_{\hat{P}_{\mathbf{k}}}=\sum_{n=1}^{N}S(\sigma^{k_n}_n),
\end{equation}
where $S(I)=0$, $S(X)=S(Y)=S(Z)=1$.  Equivalently, $l_{\hat{P}_{\mathbf{k}}}$ can be obtained by counting the number of non-zeros in the quaternary number $\mathbf{k}=k_1k_2...k_N$.


\subsection{Bell measurements}
We first introduce some interesting properties of Bell states related to Pauli operators. Then, a scheme for measuring the operator size is given.

The Bell states are two-qubit maximally entangled states. There are four Bell states~(Bell basis), which are given as follows,
\begin{equation}\label{BellStPpt} 
\begin{split}
|B^0_n\rangle=&\frac{1}{\sqrt{2}}\left(|0_n0_{n'}\rangle+|1_n1_{n'}\rangle\right)\\
|B^x_n\rangle=&X_n|B^0_n\rangle=\frac{1}{\sqrt{2}}\left(|1_n0_{n'}\rangle+|0_n1_{n'}\rangle\right)\\
|B^y_n\rangle=&Y_n|B^0_n\rangle=\frac{i}{\sqrt{2}}\left(|1_n0_{n'}\rangle-|0_n1_{n'}\rangle\right)\\
|B^z_n\rangle=&Z_n|B^0_n\rangle=\frac{1}{\sqrt{2}}\left(|0_n0_{n'}\rangle-|1_n1_{n'}\rangle\right),\\
\end{split}
\end{equation}
where $X_n,Y_n,Z_n$ is the corresponding quantum gate on the qubit $n$ and $n'$ represents the ancillary qubit. The four Bell basis are orthogonal to each other. Thus, a mapping can be established between four Pauli operators $\{I,X,Y,Z\}$, and four Bell states $\{|B^0\rangle,|B^x\rangle,|B^z\rangle,|B^z\rangle\}$, respectively. For convenient, we interchangeably use the notions $|B^x_n\rangle=|B^1_n\rangle$, $|B^y_n\rangle=|B^2_n\rangle$, $|B^z_n\rangle=|B^3_n\rangle$.

The imprinting of information of Pauli operators into Bell states can be generalized into
Pauli basis. For instance, considering a two-qubit system and an operator
\begin{equation}\label{EOP} 
	\hat{O}_f=C_0I_1I_2+C_1X_1I_2+C_4I_1X_2+C_7Z_1X_2,
\end{equation}
one can perform $\hat{O}_f$ on an initial $|B^0_1B^0_2\rangle$, which leads to
\begin{equation}\label{EFSR} 
	|\psi_f\rangle=C_0|B^0_1B^0_2\rangle+C_1|B^1_1B^0_2\rangle+C_4|B^0_1B^1_2\rangle+C_7|B^3_1B^1_2\rangle.
\end{equation}
Now, each Pauli basis is mapped into a product of Bell basis~(General Bell basis). The operator size can be obtained by counting the number of Bell basis that is not $|B^0\rangle$ in the general Bell basis, which can be implemented with projective measurements. One can evaluate the average operator size as,
\begin{equation}\label{EOPSR} 
	L[\hat{O}_f]=|C_0|^2\times0+\left(|C_1|^2+|C_4|^2\right)\times1+|C_7|^2\times2.
\end{equation}
A generalization to $N$-qubit system can be straightforward.

\begin{figure}[htbp] \centering
	\includegraphics[width=8.8cm]{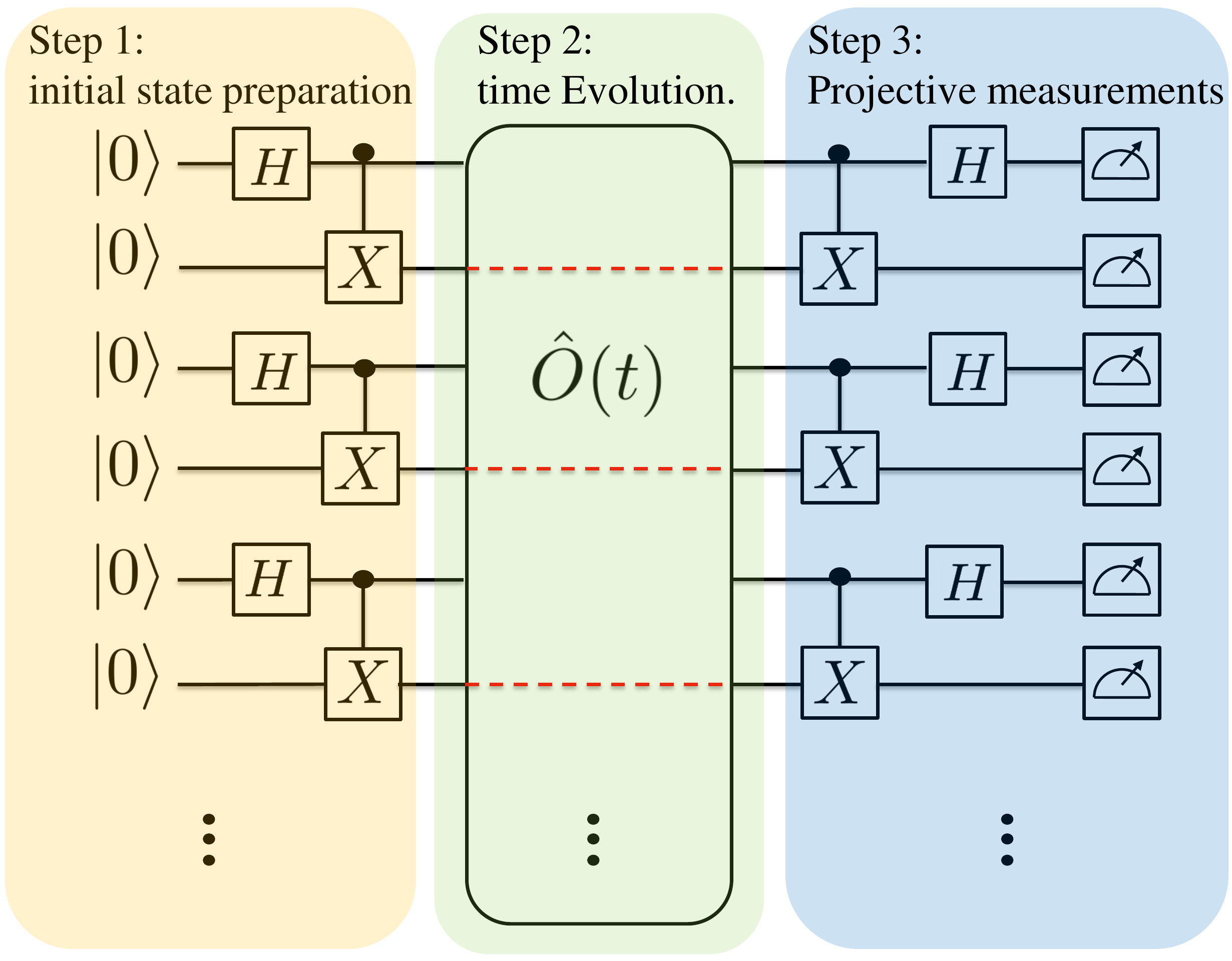}
	\caption{(Color online). The schematic diagram for the quantum circuit to calculate the operator size of operator $\hat{O}(t)$, where $H$ is the Hadamard gate, $X$ is the $X$ gate, and $\hat{O}(t)$ is the operator.}\label{p1}
\end{figure}

Now we present a procedure to evaluate the operator size with Bell measurements, which consists of three main steps~(also illustrated in Fig.~\ref{p1}).

\emph{Step 1: initial state preparation.}
Prepare the initial state as a product of $N$ Bell states,
\begin{equation}\label{ISC} 
	|\psi_0\rangle=\bigotimes^N_{n=1}|B^0_n\rangle.
\end{equation}
For a $N$-qubit quantum system, a total $2N$ qubits is required.

\emph{Step 2: time Evolution.} Perform the unitary operator $\hat{O}(t)=e^{i\hat{H}t}\hat{O}(0)e^{-i\hat{H}t}$ on the initial state $|\psi_0\rangle$, and one gets
\begin{equation}\label{FSC} 
	\begin{split}
		|\psi(t)\rangle=&\hat{O}(t)|\psi_0\rangle\\
		=&\sum_{\mathbf{k}=0}^{4^N-1}C_{\mathbf{k}}(t)\hat{P}_{\mathbf{k}}|\psi_0\rangle\\
		=&\sum_{\mathbf{k}=0}^{4^N-1}C_{\mathbf{k}}(t)\bigotimes^N_{n=1}|B^{k_n}_n\rangle\\
		=&\sum_{\mathbf{k}=0}^{4^N-1}C_{\mathbf{k}}(t)|\mathbf{k}\rangle,\\
	\end{split}
\end{equation}
where we have denoted $|\mathbf{k}\rangle\equiv\bigotimes^N_{n=1}|B^{k_n}_n\rangle$.

\emph{Step 3: Projective measurements.}
Perform the projective measurement,
\begin{equation}\label{BMM} 
	\hat{M}=\sum^N_{n=1}\hat{M}_n,~~\hat{M}_n=1-|B^0_n\rangle\langle B^0_n|.
\end{equation}
The expectation value of $\hat{M}$ is equal to the averaged operator size $L[\hat{O}(t)]$, which is derived from the following,
\begin{equation}\label{BMMR} 
	\begin{split}
		\langle\psi(t)|\hat{M}|\psi(t)\rangle=&\sum_{\mathbf{k}=0}^{4^N-1}\left|C_{\mathbf{k}}(t)\right|^2\langle\mathbf{k}|\hat{M}|\mathbf{k}\rangle\\
		=&\sum_{\mathbf{k}=0}^{4^N-1}\left|C_{\mathbf{k}}(t)\right|^2\sum^N_{n=1}S(k_n)\\
		=&\sum_{\mathbf{k}=0}^{4^N-1}\left|C_{\mathbf{k}}(t)\right|^2l_{\hat{p}_{\mathbf{k}}} =L[\hat{O}(t)]
	\end{split}
\end{equation}
Here $S(k_n)=\langle\mathbf{k}|\hat{M}_n|\mathbf{k}\rangle$ with $S(0)=0, S(1)=S(2)=S(3)=1$. As $S(k_n)=S(\sigma^{k_n}_n)$, we have $\sum^N_{n=1}S(k_n)=\sum_{n=1}^{N}S(\sigma^{k_n}_n)=l_{\hat{p}_{\mathbf{k}}}$.


The procedure of initial state preparation, time evolution, and projective measurement should be repeated to evaluate $L[\hat{O}(t)]$ with an acceptable statistical error.
Some remarks are in order. Firstly, the Hamiltonian evolution $e^{-i\hat{H}t}$ for the Hamiltonian $H$ and its reversion $e^{i\hat{H}t}$ may be implemented directly on an analog quantum simulator by engineering the Hamiltonian $H$ and $-H$, respectively. For a digital quantum computer, the evolution should be decomposed into sequences of quantum gates by Trotterization. Here we focus on the digital quantum simulation. Secondly, by performing measurement $\hat{M}$, one can also access the distribution of operator size, which takes integer values $M=0,1,2,...,N$. Thirdly, the projective measurement can be decomposed as,
\begin{equation}\label{BMMR2} 
\sum^N_{n=1}\langle\psi(t)|\hat{M}_n|\psi(t)\rangle=\sum^N_{n=1}L_n[\hat{O}(t)],
\end{equation}
where $L_n[\hat{O}(t)]$ can be taken as the operator density at site $n$ for the operator $\hat{O}(t)$. Thus, the averaged operator size is a summation of local operator densities.

\section{Simulation results}
\label{s3}
In this section, we demonstrate the quantum algorithm using a model Hamiltonian with numeral simulations. We consider both the Trotterization error and quantum noises which are two necessary ingredients when implementing the quantum algorithm. The numeral simulation is conducted using the open-source package \emph{qibo}.

As an example, we consider the mixed field Ising model(MFIM), which is a typical quantum chaotic system. The Hamiltonian of MFTM reads,
\begin{equation}\label{MFIM}
\hat{H}_{I}=\sum_{n=1}^{N-1}JZ_{n}Z_{n+1}+h_x\sum_{n=1}^NX_n+h_z\sum_{n=1}^NZ_n.
\end{equation}

The MFTM reduces to the transverse field Ising model(TFIM) at $h_z=0$, which is an integrable system. In the demonstration, the system size is taken as $N=5$ and the operator investigated is $X_3(t)$. This takes $10$ qubits in the quantum computation as one Bell state needs two qubits.

\begin{figure}[htbp] \centering
\includegraphics[width=9cm]{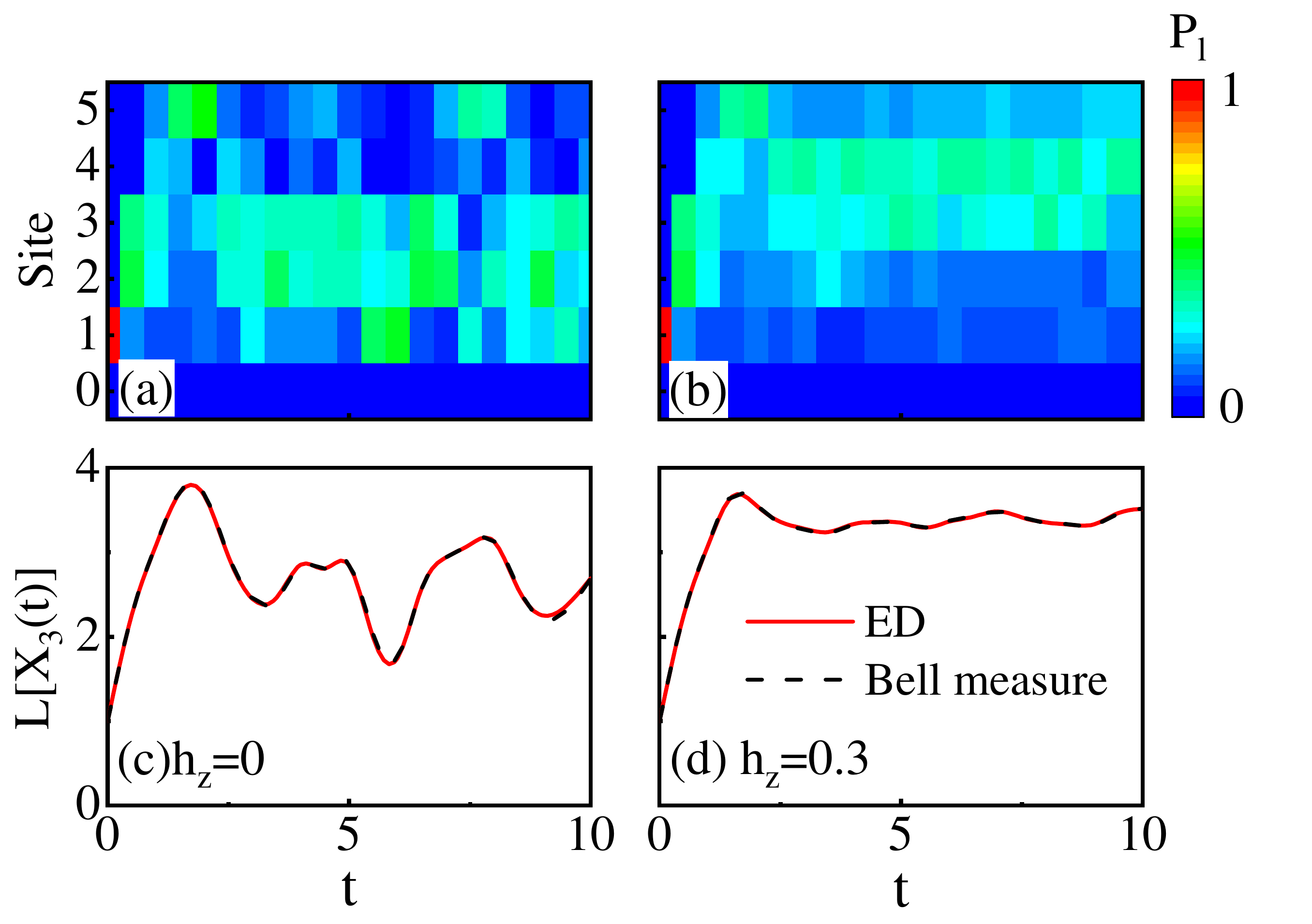}
\caption{(Color online). The operator size of $X_3(t)$ for the quantum Ising model with system site $N=5$. (a) The operator size distribution of the integrable system($h_z=0$) with different time $t$. (b) The quantum chaotic one($h_z=0.3$). Both of them are numerical results by using quantum computer simulator without circuit noise. (c, d) The corresponding operator size for (a, b). The red lines are ED results and the
break dashed lines are the result of quantum computer simulator by Bell Measurements without circuit noise.}\label{p2}
\end{figure}

In the ideal situation where both Trotter error and quantum noises are ignored, the results of operator size growth for both MFTM~($h_z\neq0$) and TFIM~($h_z=0$) are shown in Fig.~\ref{p2}. For $h_z=0$, the operator size shows an oscillation with time, which characterizes an integrable system. For $h_z\neq0$~(we chose $h_z=0.3$), the operator size increases for a period and reaches saturation afterward. In addition, the operator size distributions and their evolution with time are presented in Fig~\ref{p2}(a) and (b), respectively. Note that the simulation results fit well with the exact diagonalization(ED) for the operator size growth, which verifies the quantum algorithm.

\begin{figure}[htbp] \centering
\includegraphics[width=8.5cm]{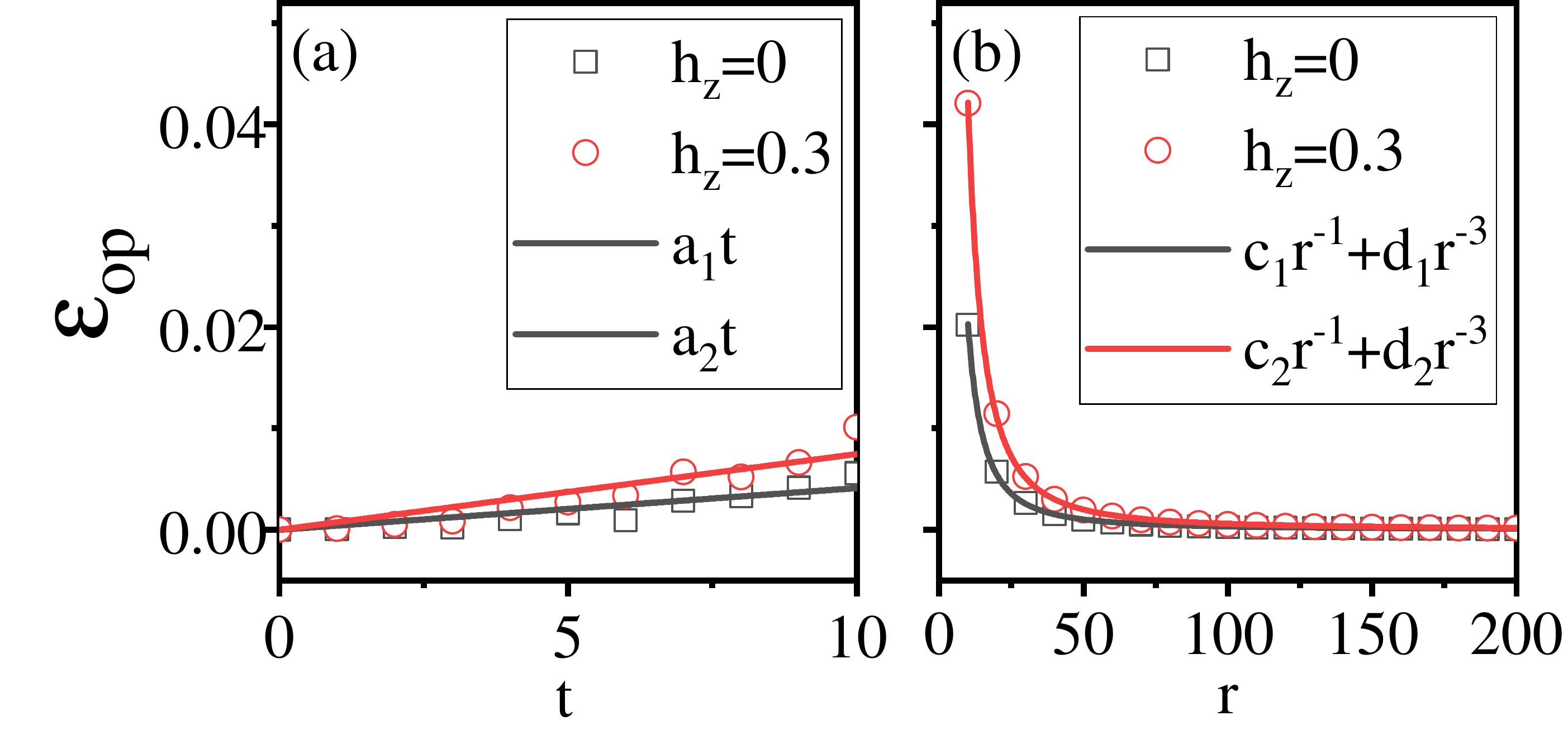}
\caption{(Color online). (a) The Trotter errors with time for both integrable system($h_z=0$) and quantum chaotic one($h_z=0.3$) with the Trotter step $r=100$. (b) The Trotter errors with different Trotter steps for a total time $t=2$.}\label{p3}
\end{figure}

We now consider the Trotter errors. To implement the Hamiltonian evolution in $\hat{O}(t)$ on a digital quantum computer, a decomposition into sequences of one-qubit and two-qubit gates is necessary, and the Trotter error $\varepsilon$ due to the decomposition is related to the evolution time $t$ and the number of time slices~(or Trotter steps) $r=t/dt$ used in the decomposition. For a generic quantum system, the relation is $\varepsilon=O(t^2/r)$. The Trotter error can be smaller for specific Hamiltonian. For instance, for a Hamiltonian $H$ with a partition into $H=H_1+H_2$ that all terms in $H_1$~(or $H_2$) mutually commutes, the Trotter error becomes, $\varepsilon= O(\frac{nt}{r}+\frac{nt^3}{r^2})$~\cite{MCTran2020}. This is just the case for MFTM or TFIM, where the Hamiltonian can be written as $H_I=H_{z}+H_{x}$ with $H_z$ and $H_x$ consisting of $Z$ and $X$ operators, respectively. The Trotter error $\varepsilon_{op}$ for the operator $\sigma_{\alpha}^k(t)=U^{\dag}(t)\sigma_{\alpha}^{k}(0)U(t)$ is $\varepsilon_{op}\sim \varepsilon$, which turns to be,
\begin{equation}\label{Ter} 
\varepsilon_{op}\sim O(\frac{t}{r}+\frac{t^3}{r^2}).
\end{equation}
Notably, the Trotter error is demonstrated by $O(\frac{t}{r})$ for small $t$.
To verify the behavior of Trotter error, we consider a maximum evolution time $T=10$. The number of Trotter steps is fixed $r=100$. By simulation, the Trotter errors with time are shown in Fig.~\ref{p3}(a)(The fitting function is marked), which are in agreement with the theoretical analysis in~Eq.(\ref{Ter}). In addition,  behaviors of Trotter errors with the number of time slices are shown in Fig.~\ref{p3}(b) for $t=2$~(at $t=2$ the operator size ceases to increase as seen in Fig.~\ref{p2}(c,d)). Again, the simulation results are in agreement with the theoretical analysis in~Eq.(\ref{Ter})(The fitting function is marked in Fig.~\ref{p3}(b)).

To implement the quantum algorithm on near-term quantum processors, the effects of quantum noises cannot be ignored~\cite{JPreskill2018}. For this, we continue to include quantum noises in the numeral simulation. As for a demonstration, we chose a noise model with depolarizing noises. The depolarizing noise model can be effective for large-size quantum circuits. During the simulation, the number of Trotter steps is set as $r=t/0.1$ and the noises of depolarization are added to each qubit after a quantum gate with a noise rate $p$. For a single qubit, the depolarization will evolve a density matrix $\rho$ to $\rho'$ as,
\begin{equation}\label{DPNE}
\rho'=\frac{p}{3}(X\rho X+Y\rho Y+Z\rho Z)+(1-p)\rho.
\end{equation}
To see the effects of depolarizing noise on the operator size, it is useful to consider a quantum operation of depolarizing on one qubit of the Bell state $|B^0\rangle$. The final state will be a mixed state of all four Bell states, $\frac{p}{3}(B^1+B^2+B^3)+(1-p)B^0$, where we have short-noted $B^k=|B^k\rangle\langle B^k|$.
Similarly, for Bell state $|B^1\rangle$, the mixed state after depolarization becomes
$\frac{p}{3}(B^0+B^2+B^3)+(1-p)B^1$. Corresponding to Pauli operators, it can be seen that the probability of turning an identify $I$ into Pauli operators $\{X,Y,Z\}$ is $p$, while the probability of turning a Pauli operator into $I$ is $\frac{p}{3}$. In this regard, depolarizing noises tend to increase the local operator density and thus the operator size.

\begin{figure}[htbp] \centering
\includegraphics[width=8.9cm]{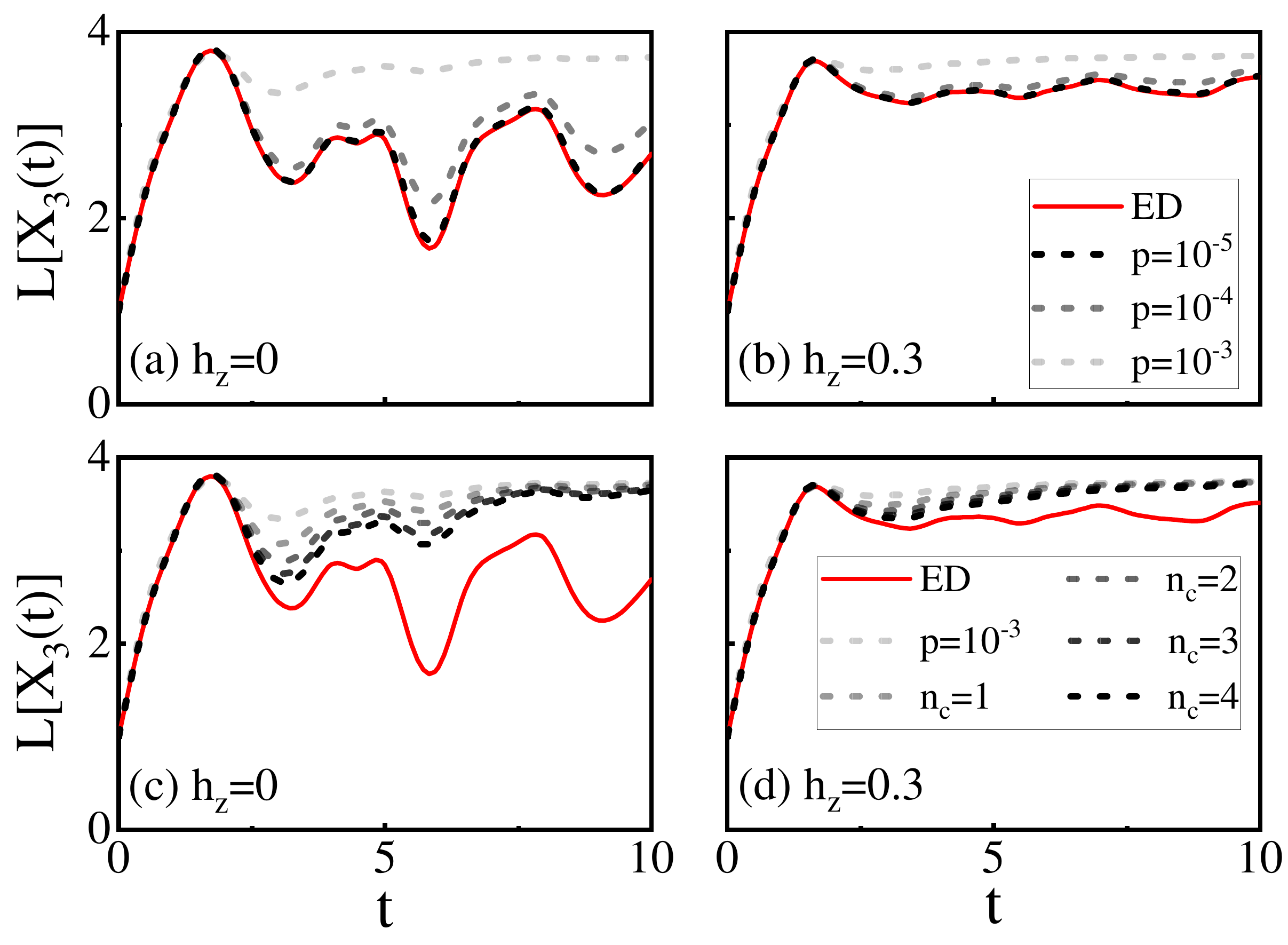}
\caption{(Color online). Noise effects and error mitigation for the operator size growth. (a,b) The operator size of  $X_3(t)$ for the integrable system and chaotic one with different depolarizing noise rates. (c,d) Results of error mitigation for (a,b) respectively by extrapolation to the zero noise limit with different cancels powers $n_c$ for a noise rate $p=10^{-3}$.}\label{p4}
\end{figure}

As shown in Fig.~\ref{p4}(a)(b), the distinction of the behavior of operator size growths can be still identified between integrable and chaotic systems when the depolarizing noise rate is small, e.g., $p\leq10^{-4}$. However, for the larger noise rate $p\sim 10^{-3}$, which is comparable to real quantum hardware, the operator size reaches saturation for both integrable and chaotic systems. As analyzed before, the depolarizing noise can also increase the operator size, and it is hard to tell integrable and chaotic systems apart even when the noise rate is large compared to NISQ quantum devices. Thus, to simulate quantum chaotic systems by studying the operator size growth on real quantum processors, it is demanding to correct the errors or reduce the effects of noises.

While quantum error correction can provide an ultimate solution for handling quantum noises~\cite{PWShor1995,AMSteane1996,DGottesman2009,DanielFVJames2001} and there are rapid progresses in experiments recently. Due to limited quantum resources, a practical way is to reduce noise effects by error mitigation techniques. We adopt an error-mitigation technique by extrapolating to the zero noise limit with Richardson's deferred method~\cite{KristanTemme2017}(see Appendix~\ref{QEM}), which is suitable for short-depth quantum circuits.
This can be achieved by performing quantum computing at several different noise rates with a scaling $p_j=c_jp$ where $p$ is the lowest noise rate feasible on the quantum hardware and $c_j$ is the scaling factor, and then make a linear combination of results with a weighting $\gamma_j$ for the noise rate $p_j$ and make extrapolation to the zero-noise limit. The parameters $c_j$ and $\gamma_j$ should satisfy two relations
\begin{equation}\label{MTCDs} 
\begin{split}
\sum_{j=0}^{n_c}\gamma_j=&1\\
\sum_{j=0}^{n_c}(\gamma_jc^k_j)=&0,\quad \text{for} \quad k=1,2,..,n_c,\\
\end{split}
\end{equation}
the error can be reduced to $O(p^{n_c+1})$, where $n_c$ is the cancels power for mitigating error. We chose parameters $c_j$ and $\gamma_j$ for different $n_c$: $c_0=1, c_1=2, \gamma_0=2, \gamma_1=-1$ for $n_c=1$; $c_0=1, c_1=2, c_2=3, \gamma_0=3, \gamma_1=-3, \gamma_2=1$ for $n_c=2$; $c_0=1, c_1=2, c_2=3, c_3=4, \gamma_0=4, \gamma_1=-6, \gamma_2=4, \gamma_3=-1$ for $n_c=3$; $c_0=1, c_1=2, c_2=3, c_3=4, c_4=5, \gamma_0=5, \gamma_1=-10, \gamma_2=10, \gamma_3=-5, \gamma_4=1$ for $n_c=4$. As seen in Fig.~\ref{p4}(c) and (d), simulation results after error mitigation get better as cancels powers $n_c$ increase when the evolution time is not too long, which is expected as the corresponding quantum circuit depth is short.

Notably, the oscillation behavior of operator size growth for the integrable system, which is lost at a noise rate $p=10^{-3}$ , restores after error mitigation  with cancels powers $n_c=3, 4$. This suggests that error mitigation can be an important ingredient for simulating and characterizing quantum integrable and chaotic systems on near-term quantum devices.

\section{conclusion}
\label{s4}
In summary, we have proposed a quantum algorithm to evaluate the operator size for quantum systems, which can be useful to characterize quantum chaotic systems on near-term quantum devices by investigating the operator size growth. By preparing a product of Bell states as the initial state, the information of the operator in terms of Pauli basis will be revealed in the Bell basis, and the operator size and its distribution can be extracted with Bell measurements. For implementing the quantum algorithm, we have considered both the Trotter errors due to the decomposition of the Hamiltonian evolution and the effects of quantum noises. We have demonstrated with numeral simulations that error mitigation is necessary for telling quantum integrable and chaotic systems apart on noisy quantum devices. Our work has suggested a feasible scheme for studying quantum chaotic systems on near-term quantum computers by measuring the operator size growth.

\emph{Acknowledgements.}---The authors thank Dr. Zhi Li for helpful suggestions. This work was supported by the National Natural Science Foundation of China (Grant No.12005065) and the Guangdong Basic and Applied Basic Research Fund
(Grant No.2021A1515010317).


\begin{appendix}
\section{Quantum error mitigation}
\label{QEM}
Following Ref.~\cite{KristanTemme2017}, we give a short description of error mitigation by zero-noise extrapolation with Richardson's deferred method.
For noisy-free system, the density matrix $\rho_0(T)$ satisfies the motion equation
\begin{equation}\label{DMMEQ}  
-i\hbar\frac{\partial\rho_0(T)}{\partial t}= [\hat{H},\rho_0(T)],
\end{equation}
which determines $\rho_0(T)=\rho_0(0)+\frac{i}{\hbar}\int^T_0[\hat{H},\rho_0(t)]dt$. The expectation value of operator $\hat{A}$ can be read as $E_0(T)=\text{tr}(\hat{A}\rho_0(T))$.

When the noisy errors are introduced to the system with a noise rate $p$, the expectation value of operator $\hat{A}$ can be rewritten as
\begin{equation}
E_{p}(T)=\text{tr}(\hat{A}\rho_0(T))+\sum^n_{k=1}a_kp^k+R_{n+1}(p,T),
\end{equation}
where $a_k$ is the coefficients of the $p^k$ and $R_{n+1}(p,T)$ is the higher order $p$ error. By considering the noises scaling $p_j=c_jp$, one can get
\begin{equation}
E_{p_j}(T)=\text{tr}(\hat{A}\rho_0(T))+\sum^n_{k=1}a_kc_j^kp^k+R_{n+1}(c_jp,T),
\end{equation}

Thus, one can obtain the mitigation result by calculation
\begin{equation}\label{MTC} 
\begin{split}
E^{n_c}_{p}(T)=&\sum_{j=0}^{n_c}(\gamma_jE_{p_j}(T))\\
=&\sum_{j=0}^{n_c}\gamma_j\left[E_0(T)+\sum^{n_c}_{k=1}a_kc^k_jp^k+R_{{n_c}+1}(c_jp,T)\right]\\
=&\sum_{j=0}^{n_c}\gamma_jE_0(T)+\sum_{j=0}^{n_c}\gamma_j\sum^{n_c}_{k=1}a_kc^k_jp^k\\
&+\sum_{j=0}^{n_c}\gamma_jR_{{n_c}+1}(c_jp,T)\\
=&\left(\sum_{j=0}^{n_c}\gamma_j\right)E_0(T)+\sum^{n_c}_{k=1}\left(\sum_{j=0}^{n_c}\gamma_jc^k_j\right)a_kp^k\\
&+\sum_{j=0}^{n_c}\gamma_jR_{n+1}(c_jp,T).\\
\end{split}
\end{equation}
The above equation Eq.~(\ref{MTC}) shows that the noisy error can be reduce to $\sum_{j=0}^{n_c}\gamma_jR_{{n_c}+1}(c_jp,T)$, by selecting a set of suitable $c_j$ and $\gamma_j$. Both $c_j$ and $\gamma_j$ should satisfy
\begin{equation}\label{MTCD} 
\begin{split}
\sum_{j=0}^{n_c}\gamma_j=&1\\
\sum_{j=0}^{n_c}(\gamma_jc^k_j)=&0,\quad \text{for} \quad k=1,2,..,n_c\\
\end{split}
\end{equation}
Then, the mitigation result can be rewritten as
\begin{equation}\label{MTCR} 
\begin{split}
E^{n_c}_{p}(T)=E_0(T)+\sum_{j=0}^{n_c}\gamma_jR_{{n_c}+1}(c_jp,T),
\end{split}
\end{equation}
where $n_c$ can be realized as cancels powers and the optimization result shows that the error reduces to $\sum_{j=0}^{n_c}\gamma_jR_{{n_c}+1}(c_jp,T)$.

\end{appendix}

\bibliography{Ref}

\end{document}